\documentclass[letterpaper,11pt]{article}

\usepackage{amsmath}
\usepackage{amsfonts}
\usepackage{amssymb}
\usepackage{graphicx}
\usepackage{color}

\usepackage{graphicx}
\usepackage[body={6.5in, 9in}, right=1in, top=1in]{geometry}
\usepackage{amssymb}
\usepackage{amsmath} 
\usepackage[numbers,sort&compress]{natbib}

\def\be{\begin{equation}}
\def\ee{\end{equation}}

\newcommand\eea{\end{eqnarray}}
\newcommand\bea{\begin{eqnarray}}
\newcommand{\sfrac}[2]{{\textstyle\frac{#1}{#2}}}

\newcommand{\Mpl}{M_{\textrm{Pl}}}

\newcommand\di{\partial}

\newcommand{\nn}{\nonumber}
\newcommand{\half}{\frac{1}{2}}
\def\({\left(}
\def\){\right)}

\newcommand{\Comment}[1]{{}}
\definecolor{MyDarkBlue}{rgb}{0.15,0.15,0.45}
\usepackage[linktocpage=true]{hyperref}
\hypersetup{
colorlinks=true,
citecolor=MyDarkBlue,
linkcolor=MyDarkBlue,
urlcolor=MyDarkBlue,
}

\begin{document}
\def\thefootnote{\fnsymbol{footnote}}

\begin{center}
\Large{\textbf{Subluminal Galilean Genesis}} \\[0.5cm]
 
\large{Paolo Creminelli$^{\rm a}$, Kurt Hinterbichler$^{\rm b,c}$, Justin Khoury$^{\rm b}$,\\[.1cm]
Alberto Nicolis$^{\rm d}$, and Enrico Trincherini$^{\rm e,f}$}
\\[0.5cm]

\small{
\textit{$^{\rm a}$ Abdus Salam International Centre for Theoretical Physics\\ Strada Costiera 11, 34151, Trieste, Italy}}

\vspace{.2cm}

\small{
\textit{$^{\rm b}$ Center for Particle Cosmology, Department of Physics and Astronomy, \\ University of Pennsylvania, Philadelphia, PA 19104, USA}}

\vspace{.2cm}

\small{
\textit{$^{\rm c}$ Perimeter Institute for Theoretical Physics, \\ 31 Caroline St. N, Waterloo, Ontario, Canada, N2L 2Y5}}

\vspace{.2cm}

\small{
\textit{$^{\rm d}$ Department of Physics and ISCAP, \\ Columbia University, New York, NY 10027, USA}}

\vspace{.2cm}

\small{
\textit{$^{\rm e}$ Scuola Normale Superiore, piazza dei Cavalieri 7, 56126, Pisa, Italy}}

\vspace{.2cm}

\small{
\textit{$^{\rm f}$ INFN - Sezione di Pisa, 56100 Pisa, Italy}}

\vspace{.2cm}

\end{center}

\vspace{.8cm}

\hrule \vspace{0.3cm}
\noindent \small{\textbf{Abstract}\\
We put forward an improved version of the Galilean Genesis model that addresses the problem of superluminality. We demote the full conformal group to Poincar\'e symmetry plus dilations, supplemented with {\em approximate} galilean shift invariance in the UV and at small field values. In this way fluctuations around the NEC-violating cosmological background are made substantially subluminal, and superluminality cannot be reached by any small change of the solution, in contrast with the original model. Dilation invariance still protects the scale-invariance of correlation functions of a massless test scalar --- which is the source of the observed cosmological fluctuations --- but the explicit breaking of the conformal group can be potentially observed in higher-order correlators. We also highlight a subtlety in matching the NEC-violating phase with the standard cosmological evolution, and discuss the possible couplings of the Galileon to gravity.
} 
\vspace{0.3cm}
\noindent
\hrule
\def\thefootnote{\arabic{footnote}}
\setcounter{footnote}{0}

\section{Introduction}

There is a recurrent yet somewhat vague connection between the Null Energy Condition (NEC) for the stress-energy tensor and the subluminality of signal propagation. {\em (i)}~The NEC is a close relative of the dominant energy condition, which states that for any observer the energy-momentum density should be a time-like future directed vector --- that is, no energy should leave the observer's future lightcone. {\em (ii)}~In general relativity the NEC implies that no closed time-like curve can form, provided suitable boundary conditions are satisfied \cite{Hawking:1991nk}. {\em (iii)}~Perhaps related to this, for linearized gravity in asymptotically flat spacetime, the NEC ensures that the light-cone defined by the gravitational field is always narrower than that defined by the underlying Minkowski spacetime \cite{VBL, AADNR}.
{\em (iv)}~For effective theories involving scalar fields only, at lowest derivative level the subluminality of excitations about a given solution implies that the solution obeys the NEC \cite{Dubovsky:2005xd}.
{\em (v)}~For the conformal Galileon~\cite{Nicolis:2008in}, certain solutions violate the NEC, others feature superluminal excitations \cite{NRT, CNT}.
In each of these instances the connection is far from vague, yet it is of a different nature every time. Moreover, it is very indirect in the example of {\em (v)}, where the NEC-violating solutions are perfectly sensible yet the same effective theory admits other solutions (including small perturbations of the NEC-violating ones) that unavoidably exhibit superluminality. 
These facts hint at either of two opposite conclusions: (1)~That the NEC/subluminality connection is generic and much deeper than shown by each individual example (and so far we have not been able to make it explicit); (2)~That it is accidental, and peculiar to the examples above.

In this paper, we sever such a connection. We will exhibit an effective field theory (EFT) with the following properties:
\begin{enumerate}

\item 
There exists a homogenous and isotropic NEC-violating solution that is stable against small perturbations.

\item 
About this solution, perturbations are generously sub-luminal, with $\delta c/ c \sim 1$. This makes their subluminality particularly robust --- it will survive generic deformations of the background solution, as long as these are not too large. This is in sharp contrast with the NEC-violating conformal Galileon of \cite{NRT, CNT}.

\item
The structure of our effective Lagrangian is protected against large quantum corrections by approximate symmetries.

\end{enumerate}
Ideally, one would like to supplement these properties with:
\begin{itemize}
\item[4.]
A Poincar\'e invariant solution, also stable against small perturbations. 

\item[5.]
A Lorentz invariant $S$-matrix about this solution,  obeying the standard positivity constraints coming from relativistic dispersion relations.

\item[6.]
Robust subluminality about this solution,  in contrast with the NEC-violating conformal Galileon~\cite{NRT, CNT}.

\end{itemize}
That is, one would like to have a perfectly well behaved relativistic field theory that can be defined starting from a Poincar\'e invariant  vacuum state, and that features NEC-violating solutions that are also perfectly well behaved. Unfortunately, at least in our simple framework, this will turn out not to be possible: conditions 4--6 contradict conditions 1--3. 
Notice however that the absence of a Poincar\'e invariant vacuum state is not necessarily an inconsistency for a relativistic effective field theory --- and we have obvious empirical evidence for this: there are systems like solids and fluids, which {\em (a)} exist as Lorentz-breaking states in a Lorentz-invariant microscopic theory (the standard model of particle physics); {\em (b)} can be described as Lorentz-breaking classical solutions in certain relativistic effective field theories \cite{Son, ENRW}, which however {\em cannot} be consistently extrapolated to zero density and used to describe the Poincar\'e invariant vacuum of the microscopic theory. For instance, we cannot hope that the vacuum of the standard model be well described by hydrodynamical equations.

Our starting point will be the conformal Galileon \cite{NRT}. We will build upon the results and the analyses of \cite{NRT, CNT}, to which we refer the reader for details about the original model and for the notation. 
There, the reason we could not avoid superluminality was essentially one of symmetry. The high-degree of symmetry of the action --- $SO(4,2)$ --- is partially broken by the NEC-violating solution, leaving the de Sitter group $SO(4,1)$ as the residual symmetry group. This degree of residual symmetry is enough
to guarantee that in the UV --- at scales much shorter than the `curvature' scale of this de Sitter solution --- excitations travel exactly at the speed of light, {\it i.e.},~on the verge of superluminality. Then, it is a matter of details whether small deformations of the background solution will admit superluminal excitations. For the conformal Galileon with nonzero $c_3$, this is always the case, and one can check that the level of superluminality is detectable within the effective theory \cite{NRT}. Setting $c_3$ to zero does not seem to improve the situation (we analyze this possibility at the end of Appendix C). One is thus led to consider the possibility of making the excitations of the background solution {\em strictly} subluminal, which can only be achieved by trimming the residual symmetries. One could consider less symmetric background solutions, which in our case would complicate many computations, or less symmetric actions, which is the direction that we will take. Obvious symmetries to jettison are the special conformal transformations, since the remainder --- 4D Poincar\'e and dilations --- close into a subgroup of $SO(4,2)$.
In Secs.~\ref{solution} and \ref{radiative} we will show that the resulting theory obeys properties 1--3 above.

Once minimally coupled to gravity, our system can drive the Galilean Genesis phase of \cite{CNT}. As far as the background evolution and the power spectrum of perturbations are concerned, the cosmological implications of our new model are the same as for the original one. However, the models become observationally  distinguishable from each other and from inflation at the level of higher-point correlation functions. We discuss the symmetry reasons behind these facts in Sec.~\ref{perturbations}.

In Sec.~\ref{reheating} we discuss an unexpected subtlety:
in the matching of our Galilean Genesis phase to a standard radiation-dominated one --- that is, at reheating --- not all the energy built up by violating the NEC can be passed on to the subsequent phases. Even for an instantaneous reheating, as a consequence of the higher-derivative structure of our interactions, an order-one fraction of the energy gets lost --- without ever violating stress-energy conservation of course: the loss of energy is due to a singularity in the equation of state at reheating, which induces a sudden, finite redshift of the energy density.
In practice this does not change the fact that we can start with vanishing energy in Minkowski space, and generate an expanding universe full of energy whose post-genesis cosmology is the same as our universe's.
However it introduces a novel question for models that, like ours, use NEC-violations as alternatives to inflation: whether the energy and the expansion rate created while violating the NEC can be inherited by the standard cosmology that comes after.

In Sec.~\ref{Dsection} we turn to generalizations of our scenario, involving higher order Galileons and non-minimal couplings with gravity, like for instance those of the `covariant' Galileon of~\cite{Deffayet:2009wt}. We argue that all these choices are radiatively stable, and we discuss in what circumstances some of the non-minimal couplings might be improvements of the effective theory.
An exhaustive analysis of all these generalizations is beyond the scope of our paper, but for the cases that we can  analyze straightforwardly, we argue that nothing is gained by considering these more general possibilities.
A number of technical derivations are collected in the Appendices.

To conclude our introductory remarks, we should also mention the ghost condensate case~\cite{ACLM}, and explain in what sense we are improving on it. There, one can have consistent NEC violations without superluminality \cite{CLNS} --- in fact excitations are always {\em extremely} sub-luminal. 
Notice that, like in our case, the theory does not admit a Poincar\'e invariant vacuum within the same effective field theory as the ghost condensate point\footnote{Both classical solutions might be allowed, formally, but they are always separated in field space by a region with ghosts, signaling the breakdown of the effective theory.}. However, in the absence of gravity, the NEC-violating branch of the ghost condensate is {\em unstable} on large scales. This might be irrelevant in practice for cosmological applications, as long as the instability scale is larger than the Hubble radius, but it is certainly a substantial difference with our case, which features no instability whatsoever, even in the absence of gravity.


\section{The NEC-violating  solution}\label{solution}
We consider a deformation of the original Genesis Lagrangian~\cite{CNT} 
\be \label{eq:alpha}
{\cal S}_\pi = \int \! {\rm d}^4 x  \bigg[ f^2 e^{2 \pi} (\di \pi)^2 + \frac{f^3}{\Lambda^3} (\di \pi)^2  \Box \pi 
+ \frac{f^3}{2 \Lambda^3}(1+\alpha) (\di \pi)^4 \bigg] \; ,
\ee
with $\Lambda \ll f$ and $\alpha$ a new dimensionless parameter of order unity\footnote{We use the mostly plus signature, so the quadratic term in \eqref{eq:alpha} has the ghostly sign.}.  We neglect gravity for the moment.  For $\alpha =0$ we recover the minimal Lagrangian studied in \cite{CNT} which non-linearly realizes full conformal invariance. For $\alpha\not=0$ conformal invariance is explicitly broken, but dilation invariance is preserved. We will address the radiative stability of this structure in Section \ref{radiative}. For the moment, we notice that for small $\pi$ 
the theory reduces to the ordinary cubic Galileon of \cite{Luty:2003vm,Nicolis:2004qq}, and so dilation invariance is enhanced to internal Galilean invariance, $\pi \to \pi + b_\mu x^\mu$. 
This statement is radiatively stable as the renormalization of the operators is local in field space, so that the action will remain Galilean invariant near the limit, and it will enjoy there all the standard non-renormalization properties of Galilean operators \cite{Luty:2003vm,Hinterbichler:2010xn,Burrage:2010cu}: in particular, quantum corrections will not generate the higher order galilean operators, nor will they renormalize the coefficients of $(\di \pi)^2$ and $(\di \pi)^2 \Box \pi$. 

We are after a `de Sitter' solution, where $e^{2\pi}$ takes the form of the conformal factor for de Sitter space~\cite{NRT, CNT,Hinterbichler:2011qk}:
\be
\label{pidesitter}
e^{\pi_{{\rm dS}}} = -\frac{1}{H_0 t} \;, \qquad -\infty < t < 0 \; ,
\ee
with $H_0$ a constant.  For the Lagrangian \eqref{eq:alpha} we find
\be \label{H0}
H_0^2 = \frac23 \frac{1}{(1+\alpha)}\frac{\Lambda^3}{f} \; .
\ee

The gravitational stress-energy tensor associated with the action \eqref{eq:alpha}, computed from the action as $T_{\mu\nu} = -\frac{2}{\sqrt{-g}}\frac{\delta S_\pi}{\delta g^{\mu\nu}}$ with minimal coupling, is that of~\cite{CNT}, 
\bea \label{Tmn}
T_{\mu\nu}^{\rm conformal}  
& = & - f^2 e^{2 \pi}\big[ 2\di_\mu \pi \di_\nu \pi -  g_{\mu\nu} (\di \pi)^2 
		\big] \nonumber\\
&-& \frac{f^3}{\Lambda^3} \big[ 2 \, \di_\mu \pi \di_\nu \pi  \Box \pi - \big(\di_\mu \pi  \, \di_\nu ( \di \pi)^2 + \di_\nu \pi \,  \di_\mu ( \di \pi)^2 \big)
+  g_{\mu\nu} \, \di_\alpha \pi \,  \di^\alpha ( \di \pi)^2
\big] \nonumber  \\
&-& \frac{f^3}{2 \Lambda^3} \big[ 4 ( \di \pi)^2 \di_\mu \pi \di_\nu \pi - g_{\mu\nu} ( \di \pi)^4 
		\big] \; ,
\eea
supplemented with
\be
\Delta T_{\mu\nu} = \alpha \frac{f^3}{\Lambda^3} \big[ -2(\di \pi)^2 \, \di_\mu \pi \, \di_\nu \pi + \sfrac12 g_{\mu\nu} (\di \pi)^4 \big] \; .
\ee
As we discuss in Section \ref{Dsection} and Appendix \ref{Dappendix}, there are ambiguities in how the higher-order Galileons are to be coupled to gravity which can affect the stress tensor, but these ambiguities do not enter here.

For a time-dependent solution the total energy density and pressure are
\begin{align}
\rho & = - f^2 \left[ e^{2\pi} \dot \pi^2 - \frac{3(1+ \alpha)}{2}\frac{f}{\Lambda^3} \dot \pi^4  \right] \;,   \label{rho}\\
p & =  - f^2 \left[ e^{2\pi} \dot \pi^2 - \frac{(1+\alpha)}{2}\frac{f}{\Lambda^3} \dot \pi^4 + \frac{2}{3} \frac{f}{\Lambda^3}\frac{\rm d}{{\rm d} t} \(\dot \pi^3 \) \right] \;,
\label{p}\end{align}
where we neglected the effect of gravity on the background solution, a good approximation at early times. 

A time-dependent solution violates the NEC if and only if the combination $\rho+p$ is negative. For the de Sitter solution \eqref{H0} we find
\be
\rho + p = - \frac{2 f^2}{H_0^2 t^4} \frac{3+\alpha}{3 (1+\alpha)} \; ,
\ee
which is negative for\footnote{Note that for the conformal case $\alpha=0$, the sign of $\rho + p$ depends only on the sign of the kinetic term, and is not affected by the cubic Galileon terms.  As we discuss in Appendix \ref{Dappendix}, this property persists for the higher Galileons with the right gravitational couplings.}
\be\label{necviolationcond}
\alpha > -1 \quad \mbox{or} \quad  \alpha < -3 \qquad \mbox{(NEC violation).}
\ee

Expanding the action to quadratic order about the de Sitter solution \eqref{H0}, we get the free action for the perturbations $\varphi \equiv \pi - \pi_{\rm dS}$,
\be \label{quadratic}
{\cal L}_{\rm quad}= \frac{f^2}{H_0^2 t^2} \left[\dot{\varphi}^2 - \frac{3-\alpha}{3 (1+\alpha)} (\nabla\varphi)^2 + \frac4 {t^2} \, \varphi^2\right] \;.
\ee
The overall coefficient is positive-definite. However, to ban instabilities, we have to make sure that the coefficient in front of the gradient energy, the signal propagation speed squared,
 \be
c_\varphi^2 = \frac{3-\alpha}{3 (1+\alpha)}
 \ee
is also positive. This is the case for 
\be\label{stabilitycond}
-1< \alpha < 3 \qquad \mbox{(stability).}
\ee
Finally --- and this is the novelty with respect to~\cite{CNT} --- we can consistently require that perturbations propagate {\em strictly} subluminally, $c_\varphi^2<1$. This happens for
\be\label{subluminalitycond}
\alpha > 0 \qquad \mbox{(subluminality).}
\ee

Demanding \eqref{necviolationcond}, \eqref{stabilitycond} and \eqref{subluminalitycond}, we find that for 
\be 0 < \alpha < 3 \qquad \mbox{(NEC violating, stable and subluminal)},\ee 
the system violates the NEC, is stable against small perturbations, and these propagate at subluminal speeds. For much larger values of $\alpha$ (at fixed $f$ and $H_0$), the cubic term becomes irrelevant, as it is the only one which does not scale with $\alpha$ on the solution; in this limit the violation of the NEC must be associated with an instability~\cite{Dubovsky:2005xd},
and, indeed, the system exhibits a gradient instability. If $\alpha$ is ${\cal O}(1)$ the speed of $\pi$ excitations is substantially less than unity, and, by continuity, generic perturbations of the de Sitter solution --- including non-infinitesimal ones --- cannot spoil this property and make the lightcone superluminal.

Notice that in our quick stability analysis above we have neglected the mass term in~\eqref{quadratic}. Its precise value is enforced by non-linearly realized time-translational invariance~\cite{NRT,Hinterbichler:2011qk,Hinterbichler:2012mv}, and it appears to have the wrong ({\it i.e.}, unstable) sign. Nevertheless, as shown in~\cite{NRT,Rubakov:2009np} the mass term in~\eqref{quadratic} is unambiguously {\em not} associated with an instability. The growing
mode solution represents a constant time shift of the background, indicating that the background solution is in fact a dynamical attractor. More generally, whenever the size of the mass term is the same as the time-variation rate of the background solution, the `mass' of excitations is ill-defined. For instance, we can make the mass term vanish via the field redefinition $\xi \equiv \varphi/ \dot \pi_{\rm dS}$ --- $\xi$ shifts by a constant under the spontaneously-broken time-translational invariance, which then forbids any mass term for it --- or we can flip its sign via some other time-dependent field-redefinition. This is related to the fact that an instability that develops on a time scale that is comparable to that over which the background solution itself changes by order one, cannot be unambiguously called an instability. 

\section{Radiative stability}\label{radiative}

We now discuss in more detail the radiative stability of the action \eqref{eq:alpha} and of the NEC-violating solution. The action is invariant under dilations, so that, neglecting gravity, only dilation-invariant operators will be generated. The most generic dilation-invariant operator is schematically of the form
\be
{\cal O}_{m,n} = e^{(4-m-n)\pi_c/f} \frac{\partial^m(\partial\pi_c)^n}{\Lambda^{2n+m-4}}\;,
\ee
where $m$ and $n$ are (non-negative) integers, $\pi_c$ is the canonically normalized field, $\pi_c \equiv f \pi$, and the powers of $\Lambda$ are such as to give the operator overall mass-dimension four. 
The dimensionless couplings with which operators of this form appear in the classical Lagrangian or get generated at quantum level are not necessarily of order one --- like the four derivative quartic term in \eqref{eq:alpha}, they can be suppressed or enhanced by suitable powers of $\Lambda/f$. Let us take this possibility explicitly into account, by defining the operators
\be
\label{eq:dilope}
{\cal O}_{m,n,q} = \left(\frac{\Lambda}{f}\right)^q {\cal O}_{m,n}  \; ,
\ee
where $q$ can in principle have either sign.

We now want to understand whether the operators that get induced quantum mechanically modify substantially the dynamics inferred from the classical Lagrangian \eqref{eq:alpha}. A way to assess this, is to estimate their size on the classical solution \eqref{pidesitter}, \eqref{H0}, and to compare it to that of the classical Lagrangian terms of~\eqref{eq:alpha}.
If we evaluate a generic ${\cal O}_{m,n,q}$ on the solution we get
\be
\label{eq:k}
\frac{1}{t^4}  \left(\frac{f}{\Lambda}\right)^3 \left(\frac{\Lambda}{f}\right)^{k} \; ,  \qquad k \equiv 1+ q +\frac{m}2 -\frac{n}2 \;.
\ee
The $t^{-4}$ scaling is a consequence of dilation invariance.
In the above notation, the classical Lagrangian \eqref{eq:alpha} reads
\be
{\cal L} \sim {\cal O}_{0,2,0} + {\cal O}_{1,3,0} + {\cal O}_{0,4,1}  \; ,
\ee
so that all terms have $k = 0$, and are all comparable\footnote{We are here interested in the typical size of each operator and we thus disregard that the contributions to the $\pi$ equation of motion of the operator $(\partial\pi)^2 \Box\pi$ cancel on our NEC-violating solution above.} on the solution, scaling as $1/t^4 \cdot (f/\Lambda)^3$.
We now want to show that the  only operators that get generated by loop diagrams have $k \ge 1$, and are therefore a negligible correction to the classical Lagrangian.

\begin{figure}[t]
\begin{center}
\includegraphics[width=5cm]{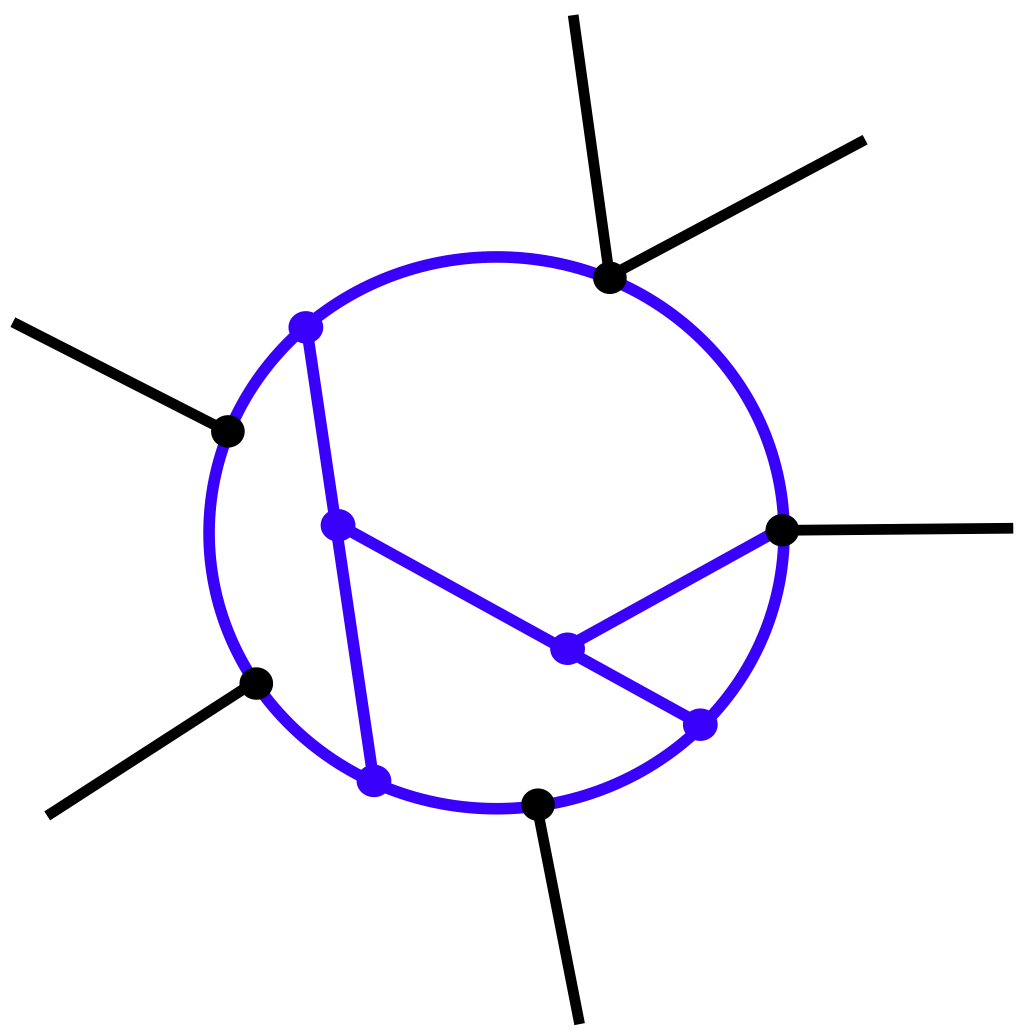}
\caption{\label{loop} \em \small Generic 1-PI loop diagram. The interaction vertices can be extracted from eq.~\eqref{eq:alpha}, upon expanding in $\pi$.}
\end{center}
\end{figure}

Consider a generic 1PI graph, like for instance that depicted in Fig.~\ref{loop}. The interaction vertices are those of~\eqref{eq:alpha}, after we expand the exponential in powers of $\pi_c$. 
For any given graph, vertices and lines can be divided into
\begin{enumerate}
\item
internal lines, and vertices that connect internal lines only (blue elements in the picture);
\item
external lines, and the vertices attached to them (black elements in the picture).
\end{enumerate}
For type-1 elements, the momenta flowing into the vertices and in the propagators are all integrated over. Assuming we cut off the loop integrals at energies of order of, or lower than, $\Lambda$ --- which is the strong coupling scale of the theory --- this part of the diagram contributes powers of $\Lambda$ times the (positive) powers of $\Lambda/f$ suppressing the interaction vertices, if they come from expanding the exponential in ${\cal O}_{0,2,0}$, or from
${\cal O}_{0,4,1}$. That is, in the parameterization of \eqref{eq:dilope}, it contributes
\be \label{type1}
m =0 \; , \quad n=0\; ,  \quad q \ge 0 \; .
\ee
For type-2 elements:
The cubic vertex ${\cal O}_{1,3,0}$ always puts two derivatives on the external leg, because of the standard non-renormalization property \cite{Luty:2003vm,Hinterbichler:2010xn,Burrage:2010cu}, so that it contributes 
\be
m =1 \; , \quad n=1\; ,  \quad q = 0 \; .
\ee 
The quartic vertex ${\cal O}_{0,4,1}$ can attach to one or two external legs; it contributes 
\be
m =0 \; , \quad n = 1, 2 \; ,  \quad q = 1 \; .
\ee
Finally, the vertices we get by expanding the  exponential in ${\cal O}_{0,2,0}$ can have at most two external legs with one derivative each, and a generic number of external legs with no derivatives, each accompanied by a $1/f$. We can focus on the case in which there are no external legs without derivatives, since those will re-sum into the correct exponential structure dictated by scale invariance, which is completely fixed once we determine the derivative structure of the operator we generate (a particular example is the term without external derivatives, which will scale as $\sim \Lambda^4 e^{4 \pi_c/f}$).  We thus get
\be \label{type2.3}
m =0 \; , \quad n = 1,2 \; ,  \quad q \geq 1 \; .
\ee
We are interested in the combination $k$ of~\eqref{eq:k}. We see that all contributions \eqref{type1}$-$\eqref{type2.3} have
\be
q + \frac{m}2-\frac{n}2 \ge 0 \; ,
\ee
so that the whole diagram has an overall $k \geq 1$, {\it i.e.}, its contribution on the solution is negligible compared with the terms we started with.

Note that we are envisioning computing quantum corrections around the Poincar\'e invariant background, but this background has a ghost.  To justify this, we appeal to the fact that the quantum effective action does not depend upon which background it is computed \cite{Coleman:1973jx}.  In fact, in applications to spontaneous symmetry breaking, the true vacuum is not known {\it a priori}.  Operationally, the ghost should not cause any problems in the computation of the effective action --- the presence of a ghost in this case only changes the sign of the propagator and of the $i\epsilon$ prescription (in fact, the action can be made ghost-free with a flip in overall sign, since we are not yet considering coupling to other matter).  The infinities due to the runaway instability of the vacuum come from phase space factors, not from the amplitudes \cite{Carroll:2003st,Cline:2003gs}, and should not be a problem here.

\section{Perturbations and their symmetries}\label{perturbations}
In Galilean Genesis, a scale invariant spectrum of perturbations is generated by a light spectator scalar \cite{CNT}, which we call $\sigma$. Given the $SO(4,1)$ symmetry of the background solution \eqref{pidesitter}, its dynamics are identical to what they would be in de Sitter space, where light scalars acquire a scale invariant spectrum~\cite{Hinterbichler:2011qk,Rubakov:2009np}. This spectrum can be later converted into adiabatic perturbations through a variety of mechanisms~\cite{Wang:2012bq}. With our new action \eqref{eq:alpha}, we have broken explicitly the original $SO(4,2)$ symmetry of the model, down to Poincar\'e plus dilation invariance, which does not contain the de Sitter isometry group $SO(4,1)$ as  a subgroup. One might worry that our predictions for cosmological observables are impaired by the absence of full de Sitter symmetry.

It is easy to see, however, that dilation invariance is all that is needed to preserve the scale invariance of $\sigma$ correlation functions\footnote{To avoid confusion, we will call `dilation invariance' the symmetry under rescalings
\be
x^\mu \to \lambda x^\mu \; , \qquad e^\pi \to \lambda^{-1}   e^\pi \; ,
\ee
while we will reserve `scale invariance' for what cosmologists mean when they refer to scale-invariant correlation functions: that they depend at most logarithmically on distances.}.
Indeed for a light scalar (whose lightness can be protected by an approximate shift symmetry), dilation invariance forces the quadratic action to take the form
\be
S_\sigma \simeq \int\!{\rm d}^4x \, \frac{A}{t^2} \left[\dot\sigma^2 - c_\sigma^2 (\vec\nabla\sigma)^2\right] \;,
\ee
where $A$ and $c_\sigma^2$ are constants~\cite{Creminelli:2011mw}. Notice that, in contrast to the $SO(4,2)$-invariant case, there is no reason to expect $\sigma$ to travel on the lightcone; a dilation invariant operator of the form $(\partial\pi\partial\sigma)^2$, for example, when evaluated on the solution~\eqref{pidesitter}, only affects the  $\dot \sigma^2$ term.  The action above yields a scale invariant spectrum, in the same way as in models of inflation with speed of sound different from unity~\cite{Garriga:1999vw}. This is easy to see via symmetry arguments \cite{CNT}, as dilation invariance forces the 2-point function $\langle\sigma(t,0) \sigma(t,\vec x)\rangle$ to depend only on the ratio $|\vec x|/t$. As the field is massless, its wavefunction becomes time-independent at late times, and therefore also independent (up to logarithms) of the spatial separation. This argument extends to higher order correlation function of $\sigma$, which will all be scale-independent as in inflation. We stress that also in inflation it is the dilation symmetry $(\eta, \vec x) \to \lambda (\eta, \vec x) $ which is the origin of  the scale invariance of correlation functions.  Indeed, only translations, rotations and (approximately) dilations are good symmetries of the inflaton background\footnote{Dilation symmetry is an approximate symmetry of inflation if all quantities, apart from the scale factor, are approximately constant. This ``slow-roll" condition is related to an approximate shift symmetry of the inflaton.} \cite{Creminelli:2011mw}.

While the spectrum is fixed by dilation invariance up to the overall normalization, the reduced symmetry of the solution will show up if we look at higher order correlation functions. In the original ``luminal'' Genesis, the action and the background solution enjoy the full $SO(4,1)$ group of symmetries, so that the dynamics of $\sigma$ are endowed with this symmetry, in strict analogy with what happens to a test scalar in de Sitter \cite{Creminelli:2011mw}. In the asymptotic future, the $SO(4,1)$ group acts as the conformal group in three Euclidean dimensions, so that the correlation functions of $\sigma$ are not only scale invariant but fully conformally invariant (with $\sigma$ transforming as a primary field of dimension zero, in the massless limit). In the present case, on the other hand, the correlation functions of $\sigma$ are only invariant under rotations, translations and dilations. These are the same symmetries of correlation functions as in single-field inflation. 

These statements applied to the linearly realized symmetries. But our subluminal Genesis differs both from the luminal case and from inflation at the level of {\em non-linearly realized} symmetries. In the original (luminal) model we start from an $SO(4,2)$ symmetry spontaneously broken down to $SO(4,1)$, a special case of the general conformal scenario of \cite{Hinterbichler:2011qk}. The broken part of the symmetry is non-linearly realized on $\pi$ and the implications of this non-linear realization are studied in \cite{CJKS}. Here, on the other hand it is the broken part of the Poincar\'e group only  that is non-linearly realized:  time translations and Lorentz boosts. This will give different constraints on the correlation functions involving $\pi$. One point in common is that the two models both non-linearly realize time translations: for instance, as discussed in~\cite{NRT,Hinterbichler:2011qk} and as mentioned in Sec.~\ref{solution}, this fixes the mass of $\pi$ excitations.
Notice that the pattern of non-linearly realized symmetries is also  quite different from single-field inflation, where the inflaton itself non-linearly realizes the $SO(4,1)$ group \cite{Creminelli:2012ed,Hinterbichler:2012nm,GHN,HHK}.  The symmetry breaking pattern for each class of models is shown in the table below.

\begin{table}[h!]
\begin{center}
\vspace{0.3cm}
\begin{tabular}{|c|c|c|} \hline
Model & Symmetry & Linearly realized \\ \hline
Single-field inflation & $SO(4,1) \times $shift  & $ISO(3) \times$dilations \\ \hline
Inflation spectator & $SO(4,1)$ & $SO(4,1)$ \\ \hline
Conformal & $SO(4,2)$ & $SO(4,1)$ \\ \hline
Subluminal GG & $ISO(3,1) \times$dilations & $ISO(3) \times$dilations \\ \hline
\end{tabular}
\caption{\small Symmetries of various cosmological models. The second line refers to a test field during inflation, with negligible coupling with the inflaton \cite{Creminelli:2011mw}. The third column indicates the symmetries linearly realized on the cosmological background: in all cases it contains rotations, translations and the dilation symmetry responsible for the observed scale-invariant spectrum.}
\end{center} 
\end{table}

We conclude that, although all these models are degenerate at the level of the spectrum of perturbations, measuring higher order correlation functions would shed light on the symmetries of the system that gives rise to density perturbations, and would distinguish our subluminal Genesis from the luminal case and from inflation.  

Before closing this section, it is worth pointing out two peculiarities of our symmetry breaking pattern. First, the de Sitter solution~\eqref{pidesitter} is {\em more} symmetric than it should be, in the sense that it has some accidental symmetries that are not present in the action~\eqref{eq:alpha}. Indeed, we have been calling it `de Sitter solution' precisely because it is invariant under the de Sitter isometry group, $SO(4,1)$. This, however, is not a subgroup of the symmetry group that leaves the action~\eqref{eq:alpha} invariant, which is just the Poincar\'e group supplemented with dilations.
The dynamics of $\pi$ perturbations and of other  fields ($\sigma$) that couple to $\pi$ are invariant under the symmetries of the action, with the spontaneously broken ones realized non-linearly. Possible accidental symmetries of the solution we expand about do not translate into accidental symmetries for the dynamics of $\pi$ and $\sigma$ excitations --- hence, the analysis in this Section is unaffected by this subtlety\footnote{Notice that having enhanced `background' symmetries is not unusual. A simpler example is given by an anisotropic model with Lagrangian $L = \dot{\phi}^2 - c^2_x (\partial_x\phi)^2- c^2_y (\partial_y\phi)^2 - c^2_z (\partial_z\phi)^2$. The background solution $\phi = {\rm const.}$ is invariant under the full Poincar\'e group, even though the action is not. }.

Second, the fact that the de Sitter solution~\eqref{pidesitter} is invariant under dilations, 
\be
\pi_{\rm dS} (x) \to \pi_{\rm dS} (\lambda x)  + \log \lambda = \pi_{\rm dS} (x)  \; ,
\ee
means that dilations are realized {\em linearly} on $\pi$'s perturbations,
\be
\varphi(x) \equiv \pi(x) - \pi_{\rm dS} (x) \to \varphi(\lambda x) \; ,
\ee
even though they are realized non-linearly on the original $\pi$ field. This is quite uncommon. Usually, when we expand an action about non-trivial field configurations, we break some of the original symmetries, and for those
broken symmetries we go from linear representations to non-linear ones. Here, instead, we start with Poincar\'e transformations realized linearly and dilations realized non-linearly on $\pi$, and we end up with the opposite.
The Poincar\'e invariant solution $e^\pi =1$ breaks dilations; the non-trivial solution we consider,~\eqref{pidesitter}, breaks some of the Poincar\'e generators (Lorentz boosts, time translations), but {\em restores} dilations.


\section{Junction with standard cosmology}\label{reheating}
The Genesis phase must eventually be followed by the standard radiation dominated phase. We will not provide an explicit model for the mechanism that reheats the universe (see \cite{LevasseurPerreault:2011mw}). Rather, we will focus on general energetic considerations and highlight a peculiarity of our scenario both in its luminal and subluminal version.

Regardless of the details of how the galileon energy gets converted into radiation, one would naively expect that for a sufficiently fast transition --- much faster than the Hubble rate --- the energy density and therefore $H$ are continuous across the transition.  The logic is that the conservation of the stress-energy tensor gives
\be
\label{eq:Nick}
\dot \rho = -3 H (\rho + p)\,,
\ee
so that the overall variation of $\rho$ is small if the transition time is short compared to $H^{-1}$. Note that this argument implicitly assumes that $p$ is not parametrically larger than $\rho$, which is usually the case. In our Genesis phase, however, we have $|p| \gg \rho$, so that the argument does not go through. This is the same as saying that the rate of variation of $\rho$ is not set by $H$, as in a standard phase, but by $1/t$, which is much larger. 

One might be led to conclude that only a transition that is rapid compared to $t$ guarantees the continuity of $H$. It turns out that this conclusion is also too hasty.
The point is that, as we make the transition faster and faster, the pressure blows up. Indeed, from the explicit form of $\rho$ and $p$ given in~\eqref{rho} and~\eqref{p}, 
we see that the last term in the expression for $p$,
\be
p_{\rm sing} \equiv -\frac{4 }{9(1+\alpha)} \frac{f^2}{H_0^2} \frac {\rm d}{{\rm d}t} \dot \pi^3 \; , \label{psingmod}
\ee
 blows up as we make an instantaneous transition in which $\dot\pi$ changes by a finite amount. 
To be concrete, let us imagine that when $\pi$ reaches a certain value $\pi_*$, there is a sudden upward change in the potential energy
\be
V(\pi) = V_0 \;\theta(\pi-\pi_*) \;.
\ee
In this way, (part of) the Galileon energy is transformed into potential energy, which later can be easily converted into radiation or anything we like. In the absence of the singular contribution to the pressure, $\rho$ would be continuous across the transition, and in principle one could dial the value of $V_0$ to get all the galileon energy converted into potential energy. However, it is easy to check that the equation of motion for $\pi$ is (obviously) equivalent to \eqref{eq:Nick} and gives
\be
\dot\rho = -3 H \cdot \left(p_{\rm sing} +\mbox{finite} \right) \; ,
\ee
where `finite' stands for all contributions to $\rho+p$ other than $p_{\rm sing}$, which are finite --- though potentially discontinuous, like $V(\pi)$ --- at the transition time.
Using the Friedmann equation, $H^2 = \rho/3M_{\rm Pl}^2$, this can be rewritten as
\be
\frac{{\rm d}}{{\rm d}t} \left(H - \frac{2}{9(1+\alpha)} \frac{f^2}{ H_0^2} \frac{\dot \pi^3} {M_{\rm Pl}^2}  \right) = \mbox{finite} \,.
\ee

If we now integrate this expression in time from slightly before to slightly after the transition, we see that it is the quantity
\be
H - \frac{2}{9(1+\alpha)} \frac{f^2}{ H_0^2} \frac{\dot \pi^3} {M_{\rm Pl}^2} 
\label{matching}
\ee
that remains constant --- rather than $\rho$ or $H$ --- across an instantaneous transition. 
In Appendix~\ref{app:covjcn}, we show how this result can be obtained covariantly 
through Israel-like junction conditions. Note that $\rho = 0$ on our Galilean Genesis solution~(\ref{pidesitter}), hence $H = 0$ to zeroth order in $1/M_{\rm Pl}$. To obtain the leading non-zero contribution to $H(t)$, we can integrate $\dot{H}M_{\rm Pl}^2 = -(\rho + p)/2$ to obtain
\be
H(t) = -\frac{f^2}{H_0^2M_{\rm Pl}^2 t^3} \frac{3 + \alpha}{9(1+ \alpha)}\,.
\label{Hexplicit}
\ee
In the NEC-violating range~(\ref{necviolationcond}), this describes an expanding universe from an asymptotically static state, as it should.
Substituting~(\ref{pidesitter}) and the above expression for $H(t)$, we find that~(\ref{matching}) equals $\frac{1 + \alpha}{3+ \alpha}H$. 
Hence, assuming $\dot \pi$ vanishes after the transition, the Hubble parameters
before and after the transition are related by
\be
H_{\rm after} = \frac{1 + \alpha}{3 + \alpha} H_{\rm before}  \qquad \qquad  \mbox{(assuming $\dot \pi \to 0$ instantaneously)}\,.
\ee
  
In principle there is no reason to restrict to an instantaneous transition. However, at any given time $t$, there is a time $|t|$ before $H$ blows up, so the transition better conclude within a time at most of order $t$.
For a smooth transition with this typical time scale, all terms in $p$ are of comparable sizes, 
\be
p \sim \frac{f^2}{H_0^2} \frac{1}{t^4} \; ,
\ee
and $\rho$ undergoes an overall variation of order one,
\be
\Delta \rho \sim t \, H \, p \sim \rho \; .
\ee
This implies
\be
H_{\rm after} \sim H_{\rm before} \qquad \qquad  \mbox{(assuming $\dot \pi \to 0 \; $ in $ \; \Delta t \sim t$)}\,.
\ee

Apart from a model-dependent numerical coefficient of order one relating $H_{\rm before}$ and $H_{\rm after}$, these results do not change the qualitative conclusion that our NEC-violating phase can take an initially vanishing or negligible Hubble rate, increase it by many orders of magnitude, and then make a transition to a standard, NEC-obeying radiation-dominated cosmology. However, they do offer an unexpected twist for the whole start-the-universe-via-NEC-violation program: not only does one have to come up with consistent NEC-violating mechanisms that create energy out of nothing, but one must also make sure that the bulk of this energy can later be passed to more standard forms of matter or radiation. 

\section{\label{Dsection} The other Galileons and coupling to gravity}

The starting point of the previous Sections was the minimal conformal Galileon Lagrangian, which includes only the kinetic term and the cubic interaction together with their conformal completion. The existence of a stable NEC-violating solution in this case forces the kinetic term to have the wrong sign \cite{NRT,CNT}, and this in turn implies that the fluctuations around the Poincar\'e invariant solution $\pi=0$ are ghost-like.   
However, there are five possible conformal Galileon terms \cite{Nicolis:2008in}, and one may wonder whether the presence of higher order interactions or their dilation invariant deformations can give rise to a stable Poincar\'e invariant vacuum, without spoiling the good properties of the de Sitter solution discussed above. 

The inclusion of the higher order Galileon terms introduces an ambiguity in the choice of the action --- and as a consequence in the definition of the NEC itself --- when the coupling of the scalar to gravity is considered. The fact that non-minimal couplings to the metric can give rise to different stress-energy tensors is not surprising, the most familiar example being the ``improved" energy-momentum tensor obtained by adding the conformal coupling $-\sfrac1{12 } R \, \varphi^2$ to the minimally coupled, massless $\lambda \varphi^4$ theory. What is new in the presence of higher order Galileon interactions is that even the definition of minimal coupling is ambiguous~\cite{Khoury:2011da}. The reason is easy to understand: consider the quartic Galileon in flat space, schematically of the form $(\partial \pi)^2 (\partial^2 \pi)^2$, then commute some derivatives and integrate by parts to obtain the same structure with a different Lorentz contraction. Now let us minimally couple the two (equivalent) structures by promoting partial derivatives $\partial$ to covariant ones $\nabla$ and contracting all indices with $g_{\mu\nu}$. We can go back from the second structure to the first one by integrating by parts again but now, since the theory has higher derivatives, there can be a non-trivial commutator $[\nabla,\nabla]$ which is proportional to the Riemann tensor, thus a term of the form $R (\partial \pi)^4$ can appear\footnote{In the case of the cubic term this ambiguity is absent because this term does not have enough derivatives to generate a non-trivial commutator.}:
the two ``minimally coupled" interactions are not equivalent and they give different contributions to $T_{\mu\nu}$. We will discuss below which one among all possible minimal couplings is the most convenient for our purposes.

So far we have concentrated on the case(s) of minimal coupling, however it has been argued that this choice may not be the healthiest. Indeed, in the presence of dynamical gravity, the contribution to the equations of motion for the scalar and for the metric perturbations  given by the quartic and quintic Galileon do not remain of second order, but also contain three derivatives for any of the possible choices of minimal coupling \cite{Deffayet:2009wt}\footnote{Again, since~\cite{CNT} and the present paper focus on the simplest scenario with only the cubic conformal Galileon, their analyses are not affected by the following discussion.}. 
Operators with three derivatives are not obviously associated with extra ghost-like states --- at worst, one needs to impose {\em one} additional initial condition, which sounds like half an additional degree of freedom.  Still, it is hard to believe that they do not impair the validity of the theory in situations where they become relevant. One can avoid them altogether by adding suitable non-minimal couplings \cite{Deffayet:2009wt}. They are truly non-minimal in the sense that they do not correspond to any of the possible minimal choices discussed above, indeed one needs terms that cannot be generated by commuting covariant derivatives. These non-minimal couplings can give a different stress-energy tensor. However, the calculation of $T_{\mu\nu}$ is considerably more complicated, even restricting to the homogeneous and isotropic case, as one cannot use the tricks discussed in \cite{NRT}.  
Even these non-minimal terms are not unique, and there are different choices which yield second order equations and reduce to the Galileons on flat space \cite{Deffayet:2009mn}.  A preferred choice for the non-minimal terms can be derived using the brane construction of \cite{deRham:2010eu,Goon:2011qf,Goon:2011uw}.  We further explore this choice and the associated stress tensors in Appendix \ref{Dappendix}.

From an effective field theory (EFT) standpoint, the three-derivative terms are not pathological insofar as they can be treated as a small perturbation. The cutoff of the EFT must be lower than the scale where these higher-derivative terms become important. This scale is easy to estimate for our cosmological solution. The Lagrangian with one of the minimal couplings to gravity takes the form \cite{NRT}
\be
\frac{f^2}{H_0^2} \phi^4 F\left(\frac{\partial\phi}{\phi^2} , \frac{\nabla\nabla\phi}{\phi^3}\right) + \sfrac12 \Mpl^2 R \; ,
\ee
where $\phi \equiv H_0 e^\pi$, 
so that the quadratic action for the perturbations, 
\be
\phi = -\frac{1}{t} +\psi \; , \qquad g_{\mu\nu} = \eta_{\mu\nu}+ h_{\mu\nu} \; , 
\ee
reads schematically
\be
\frac{f^2}{H_0^2} \left[(\partial\psi)^2 +  \partial^2\psi \partial h\right] + \Mpl^2 (\partial h)^2 \;,
\ee
where we neglected mixing terms with fewer derivatives, to be discussed shortly. Once we choose canonical normalization for the diagonal terms, the mixing takes the form 
\be
\frac{f}{H_0 \Mpl} \partial^2\psi_c \partial h_c \;.
\ee
It becomes as important as the diagonal terms at an energy scale
\be
E_{\rm max} \equiv \frac{\Mpl}{f} H_0 \;,
\ee
which sets an upper limit on the UV cutoff of the EFT\footnote{The effect of the three-derivative terms is not always small as it happens on this background. We can repeat the analysis of the previous paragraphs in the case where the Galileon field is responsible for the present acceleration \cite{Nicolis:2008in} and concentrate on the spherical solution sourced by a localized object of mass $M_*$, such as a star. 
Focusing on the region outside the Vainshtein radius, the quadratic action for the perturbations 
\be
\pi= \pi_0(r) + \varphi  =  \frac{M_*}{M^2_{\rm Pl}}\frac{1}{r}+ \varphi \; , \qquad g_{\mu\nu} = \eta_{\mu\nu}+ h_{\mu\nu} \; , 
\ee
is schematically (taking $f \sim M_{\rm Pl} $)
\be 
M^2_{\rm Pl}(\partial \varphi )^2 + \frac{M^4_{\rm Pl}}{\Lambda^6}(\partial \pi_0)^3 \partial h \partial^2 \varphi + M^2_{\rm Pl}(\partial h )^2 \;.
\ee
Going to canonical normalization one finds
\be
E_{\rm max} \sim \frac{\Lambda^6}{(\partial \pi_0)^3 M^2_{\rm Pl}} \sim \frac{(\Lambda r)^6 \Mpl^4}{M_*^3}
\ee
At a distance from the source of the order of the Vainshtein radius $r_{\rm V} \sim \big( \frac{M_*}{\Mpl} \big)^{1/3} \Lambda^{-1}$ the cutoff is comparable to the inverse Schwarzschild radius, $E_{\rm max} \sim 1/ r_{\rm S}$.
This may not be problematic for the solar system, but it further limits the validity of the EFT.}.
Notice that this scale depends on $\Mpl$, so that it is parametrically different --- though not obviously smaller or larger --- than the other UV scales discussed in \cite{NRT}, such as the strong coupling scale $\Lambda \sim (f H_0^2)^{1/3}$. 
The background solution has a typical energy scale $\partial\phi/\phi \sim 1/t$ which, neglecting gravity momentarily, sets the freeze-out frequency of scalar perturbations \cite{CNT}. Therefore cosmological observables can be consistently calculated within the EFT if this scale is smaller than $E_{\rm max}$, which is the case for sufficiently early (negative) times:
\be
\label{eq:early}
|t| \gg t_0 \equiv \frac{f}{\Mpl} H_0^{-1} \;.
\ee
This is the same regime in which we can neglect the effect of gravity on the background solution \cite{CNT}. Indeed going to early times is equivalent to sending $\Mpl$ to infinity, which makes all the gravitational effects --- including the mixing --- weaker and weaker. In particular, in this regime the freeze-out of perturbations is dominated by the $\phi$ background, as assumed above. It is easy to check that also the two-derivative mixings are negligible in the same limit, scaling as $t_0/t \cdot \partial\psi_c \partial h_c$. In conclusion the minimal coupling is perfectly consistent as long as we stick to sufficiently early times, in the sense of~\eqref{eq:early}. If the transition to the standard cosmology occurs before $t_0$, the model is consistent throughout.
In fact, we could even envision that reaching $t \sim t_0$ and consequently probing energies of order $E_{\rm max}$ is what triggers this transition: new physics must be present at $E_{\rm max}$ to save the consistency of the theory, and this new physics may be responsible for draining energy out of the Galileon sector and reheating the Universe.

The conclusion of this digression is that as far as we are interested in our early-universe scenario, the minimal couplings do not give rise to instabilities below the UV cutoff of the EFT. Notice that all these choices (minimal couplings, ``covariant improvements''  \cite{Deffayet:2009wt}, etc.) are also technically natural. As shown in~\cite{Creminelli:2010qf}, starting with minimally coupled Galileon terms, we generate operators of the form $(\partial^2\pi)^n$, and terms where two derivatives are replaced by a Riemann tensor, which are subleading.
We can then use the minimally coupled Lagrangian that is obtained from a specific form of the conformal Galileon operators, the one built starting from the curvature invariants involving the conformally flat fictitious metric $e^{2\pi} \eta_{\mu\nu}$ \cite{Nicolis:2008in}. The computation of $T_{\mu\nu}$ in this case drastically simplifies \cite{NRT}. 

We can finally answer the question raised at the beginning of this Section. When all the Galileon operators are present, the sign of the kinetic term can be reversed to give a stable $\pi=0$ solution while preserving the existence and stability of the NEC-violating one. However, it inevitably exhibits superluminality. Around the Poincar\'e invariant solution fluctuations are luminal, independently of the breaking of special conformal transformation; the leading correction to the speed of propagation around weak-field deformations is given by $(\partial \pi)^2 \Box \pi$, and it is superluminal \cite{NRT}. The only possibility to avoid superluminality would be to set the coefficient of the cubic interaction to zero, but this is not compatible with the existence of a stable NEC-violating solution as manifested by the conditions found in \cite{NRT}. This conclusion holds for $\alpha = 0$ or small enough. One could hope that for a large value of $\alpha$, not only one has subluminality around the NEC violating solution, but also the Minkowski vacuum becomes healthy. Unfortunately this is not the case, as we prove in Appendix \ref{app:Minky}.

\section{Conclusions}
Subluminal Galilean Genesis is a consistent model of early cosmology that is alternative to standard inflation. The time-dependent NEC-violating solution obeys all the basic theoretical consistency requirements: it is stable, excitations around it are largely subluminal and it comes from an action whose form is protected by approximate symmetries. The same symmetries lead to the production of a scale invariant spectrum of perturbations. The predictions of the model are not completely degenerate with the ones of inflation and a distinction between the two is possible at the level of higher order correlation functions.

Various questions remain open. In our framework we could not build a theory that, besides the cosmological solution we are interested in, describes also an healthy Minkowski vacuum. It is not clear whether this is accidental or points to a subtle inconsistency implied by the violation of the NEC. To explore this issue one could consider different couplings with gravity and/or extensions of the Galilean symmetry. Another question is whether there are other theoretical consistency checks --- analogous to the usual analyticity properties of the S-matrix in Minkowski --- that can be applied to a time dependent solution. Finally it remains open the general question of whether some fundamental properties of UV-complete relativistic theories  forbid  NEC violations, even if consistent EFTs can be written down.


\subsection*{Acknowledgements}
The work of A.N. is supported by NASA ATP under contract 09-ATP09-0049  by the DOE under contract DE-FG02-92-ER40699.  K.H. is supported by funds provided by the University of Pennsylvania, and research at the Perimeter Institute is supported by the Government of Canada through Industry Canada and by the Province of Ontario through the Ministry of Research and Innovation.  The work of J.K. is supported in part by NASA ATP grant NNX11AI95G, the Alfred P. Sloan Foundation and NSF CAREER Award PHY-1145525.

\appendix

\section{\label{Dappendix}Non-minimal couplings and the higher Galileons}

Here we calculate explicitly the possible higher Galileons along with the set of non-minimal couplings preferred by brane constructions.  We then calculate the energy density and pressure around a dS like solution, and obtain a curious result showing that the NEC is violated if and only if the coefficient of the quadratic term around flat space is ghost-like.

In \cite{Deffayet:2011gz}, it is shown how to add non-minimal terms to any kind of Galileon in order to preserve the second order equations of motion.  However, the procedure is not unique -- there are many different non-minimal terms one can add which preserve second order equations and reduce to the desired flat space Galileon as the metric goes to Minkowski.  Different choices can give different energy densities and/or pressures, even around flat space.  

A way to naturally single out a particular choice of the possible non-minimal terms of \cite{Deffayet:2011gz} is via the brane construction of \cite{deRham:2010eu,Goon:2011qf,Goon:2011uw}.  When constructed in this way, the Galileons have a 5-dimensional interpretation: the scalar, in a certain limit, is the brane bending mode of a flat brane (with higher order world-volume curvature corrections) probing AdS$_5$ with the zero mode of the bulk metric turned on.   With the metric turned off, the non-linear conformal symmetry descends from the isometry group of AdS$_5$.  Following through this construction, we find the following set of non-minimal couplings for the five conformal Galileons,
\bea  
{\cal L}_1&=&-{1\over 4}\sqrt{-g}e^{4\pi} \ , \nn\\
{\cal L}_2&=&-\half \sqrt{-g}e^{2\pi}(\partial\pi)^2 \ ,\nn \\
{\cal L}_3&=&\frac{1}{2}\sqrt{-g}(\partial {\pi})^2\( [\Pi]+\frac{1}{2} (\partial {\pi})^2 \)\ ,\nn\\
{\cal L}_4&=&\frac{1}{2}\sqrt{-g}e^{-2 \pi}(\partial  \pi)^2
\(-[ \Pi]^2+[ \Pi^2]+{1\over 2}(\partial  \pi)^2[\Pi]-{1\over 2}(\partial  \pi)^4+{1\over 4}(\partial\pi)^2 R\) \ , \nn \\
{\cal L}_5&=& \half \sqrt{-g}e^{-4 \pi} (\partial  \pi)^2
\Big[[ \Pi]^3-3[ \Pi][ \Pi^2]+2[ \Pi^3]-3(\partial  \pi)^2([ \Pi]^2-[ \Pi^2])\nn\\
&&+\frac{30}{7}(\partial  \pi)^2((\partial  \pi)^2[ \Pi]-[ \pi^3])+\frac{3}{28}(\partial  \pi)^6 \nn\\
&&+6 (\partial  \pi)^2G_{\mu\nu}\(\partial^\mu\pi\partial^\nu\pi+{1\over 4}\Pi^{\mu\nu}\)+{1\over 2}(\partial\pi)^4 R\Big] \ .\label{DBIgalderivexpan}
\eea
Some explanation of the notation is in order: the metric is $g_{\mu\nu}$, the associated covariant derivative is $\nabla_\mu$, and $G_{\mu\nu}=R_{\mu\nu}-\half R g_{\mu\nu}$ is the Einstein tensor. Meanwhile, $\Pi$ is the matrix of second derivatives $\Pi_{\mu\nu}\equiv\nabla_{\mu}\nabla_\nu\pi$.  For traces of the powers of $\Pi$ we write $[\Pi^n]\equiv {\rm Tr}(\Pi^n)$, {\it e.g.} $[\Pi]=\nabla_\mu\nabla^\mu\pi$, $[\Pi^2]=\nabla_\mu\nabla_\nu\pi\nabla^\mu\nabla^\nu\pi$, where all indices are raised with $g^{\mu\nu}$.  We define the contractions of the powers of $\Pi$ with $\nabla\pi$ using the notation $[\pi^n]\equiv \nabla\pi\cdot\Pi^{n-2}\cdot\nabla\pi$, {\it e.g.}, $[\pi^2]=\nabla_\mu\pi\nabla^\mu\pi$, $[\pi^3]=\nabla_\mu\pi\nabla^\mu\nabla^\nu\pi\nabla_\nu\pi$, where all indices are raised with respect to $g^{\mu\nu}$.

The only two terms used in the present paper and in \cite{CNT} are ${\cal L}_2$ and ${\cal L}_3$, and these have no non-minimal terms or ambiguities in the minimal coupling.

Varying the Lagrangians (\ref{DBIgalderivexpan}) with respect to the metric and then specializing to time-dependent configurations $\pi=\pi(t)$ on flat space $g_{\mu\nu}=\eta_{\mu\nu}$, we find the following energy densities and pressures,

\bea &&\rho_1 = {1\over 4}e^{4\pi}\,, \ \ \ \ \ \ \ \ p_1=- {1\over 4}e^{4\pi}\,,\\
 &&\rho_2 = {1\over 2}e^{2\pi}\dot\pi^2\,, \ \ \ \  \ p_2={1\over 2}e^{2\pi}\dot\pi^2\,,\\
  &&\rho_3 = {3\over 4}\dot \pi^4\,,  \ \ \ \ \ \ \ \ \ p_3={1\over 4}\dot\pi^2\left(\dot\pi^2-4\ddot\pi\right)\,,\\
   &&\rho_4 = {3\over 4}e^{-2\pi}\dot \pi^6\,,  \ \ \ p_4={3\over 4}e^{-2\pi}\dot\pi^4\left(\dot\pi^2-2\ddot\pi\right)\,,\\
      &&\rho_5 =  {3\over 8}e^{-4\pi}\dot \pi^8\,,  \ \ \ p_5={1\over 8}e^{-4\pi}\dot\pi^6\left(5\dot\pi^2-8\ddot\pi\right)\,.
\eea

For the dS configurations of interest $\pi(t)=-\log(-H_0t)$,  given the general lagrangian ${\cal L}=\sum_{i=1}^5 c_i{\cal L}_i$ with arbitrary coefficients $c_i$, the equation of motion for flat space reduces to 
\be {c_1}+2{c_2} H_0^2+3{c_3} H_0^4+3{c_4} H_0^6+\frac{3{c_5} H_0^8}{2}=0\,,\ee
so a solution exists for any positive real valued root to this polynomial in $H_0^2$.

The total energy density and pressure on this solution are
\bea \rho&=&\sum_{i=1}^5 c_i \rho_i=c_1\frac{1}{4 H_0^4 t^4}+c_2\frac{1}{2 H_0^2 t^4}+c_3\frac{3}{4 t^4}+c_4\frac{3 H_0^2}{4 t^4}+c_5\frac{3 H_0^4}{8 t^4}\,,\\
 p&=&\sum_{i=1}^5 c_i p_i=-c_1\frac{1}{4 H_0^4 t^4}+c_2\frac{1}{2 H_0^2 t^4}-c_3\frac{3}{4 t^4}-c_4\frac{3 H_0^2}{4 t^4}-c_5\frac{3 H_0^4}{8 t^4}\,.
 \eea
Remarkably, all the terms except those from the quadratic part ${\cal L}_2$ cancel in the combination $\rho+p$,
\be \rho+p=\frac{c_2}{ H_0^2 t^4}\,,\ee
and this is negative if any only if $c_2$ is,
\be  {\rm NEC\ violation} \Leftrightarrow \rho+p<0 \Leftrightarrow c_2<0\,.\ee
Thus, no matter what higher order Galileons are included, violating the NEC on a genesis-like solution requires a wrong sign kinetic term around the Minkowski vacuum.  Note that this cancellation for the higher order terms is something special about the special non-minimal couplings in \eqref{DBIgalderivexpan} coming from the higher-dimensional brane construction, and would not be true for minimal couplings or other possible non-minimal couplings.  In particular, it is not true of the choices made in \cite{NRT}.

\section{\label{app:covjcn}Covariant Junction Conditions}

In this Appendix we derive from a covariant standpoint the junction condition to standard cosmology obtained in~(\ref{matching}). 
In the limit of an instantaneous transition, we can approximate the transition event as a space-like surface located at some $t = t_*$. 
The action for $t\leq t_*$ is given by 
\bea
\nonumber
S[t\leq t_*] &=&  \int_{t < t_*} {\rm d}^4x\sqrt{-g}\left[ \frac{M_{\rm Pl}^2}{2} R + f^2 e^{2 \pi} (\di \pi)^2 + \frac{f^3}{\Lambda^3} (\di \pi)^2  \Box \pi 
+ \frac{f^3}{2 \Lambda^3}(1+\alpha) (\di \pi)^4\right] \\ 
&+& \int_{t=t_*^-} {\rm d}^3x\sqrt{h}\left[M_{\rm Pl}^2 K   - \frac{f^3}{\Lambda^3}\left(h^{ij}\partial_i\pi\partial_j\pi{\cal L}_n\pi + \frac{1}{3}({\cal L}_n\pi)^3\right)\right] \,,
\label{action2}
\eea
where $h_{ij}$ denotes the induced metric on the $t=t_*$ boundary, $K_{ij} \equiv {\cal L}_n h_{ij}/2$ is its extrinsic curvature, and ${\cal L}_n$ denotes as usual the Lie derivative with respect to the unit time-like vector normal to the boundary. The boundary action includes the Gibbons-Hawking term and its galileon cousin~\cite{Dyer:2009yg,Agarwal:2009gy,Padilla:2012ze} necessary to have a well-defined variational principle. Similarly for $t\geq t_*$.

Stationarity with respect to variations of the metric gives Einstein's equations, together with an Israel junction condition at $t= t_*$:
\be
\Delta \left[K \delta^i_{\; j} - K^i_{\; j} + \frac{2f^3}{M_{\rm Pl}^2\Lambda^3}\left(\partial^i\pi\partial_j\pi{\cal L}_n\pi + \frac{1}{3}\delta^i_{\; j}({\cal L}_n\pi)^3\right)  \right] = 0\,.
\ee
In other words, the quantity in square brackets is conserved across the transition. Specializing to cosmology, $K^i_{\; j} = H \delta^i_{\; j}$ and $\pi = \pi(t)$, and using~(\ref{H0}),
this reduces to
\be
\Delta \left( H - \frac{2}{9(1+\alpha)} \frac{f^2}{ H_0^2} \frac{\dot \pi^3} {M_{\rm Pl}^2} \right) = 0\,,
\ee
which reproduces~(\ref{matching}).

\section{\label{app:Minky}Healthy Minkowski vacuum and higher Galileons}
In this Appendix we show that even considering all the four conformal Galilean operators (with minimal coupling with gravity) deformed by the addition of a term $(\partial\pi)^4$ proportional to $\alpha$, as in eq.~\eqref{eq:alpha}, it is {\em not} possible to have a NEC violating solution and, at the same time, a completely healthy theory around Minkowski. 

We make the following requirements on the theory:
\begin{enumerate}
\item $c_1 = 0$, to allow for a Minkowski solution.
\item $c_2 > 0$ to avoid ghosts around Minkowski.
\item $c_3 =0$, since the DGP term, $(\partial\pi)^2 \Box\pi$, always induces superluminality around weak field solutions in Minkowski, independently of the sign of the operator \cite{NRT}. 
\item $\alpha >0 $, both to close the lightcone of perturbations around the NEC violating solution and to have an healthy 2-to-2 S-matrix in Minkowski \cite{NRT}.
\item $c_4 < 0$ to avoid superluminality around weak field solutions in Minkowski. Indeed, after setting $c_3 = 0$, the leading corrections to the lightcone come from the conformal breaking $(\partial\pi)^4$ and from the quartic Galileon. Comparing the coefficients of the two operators (with $\alpha \sim 1$), it is easy to see that the quartic Galileon dominates when the weak field solution is characterized by a length scale which is much smaller than $H_0$, $\partial \gg H_0$. Focussing on this regime, we consider the linear equations of motion around a classical background $\pi_0$ and we are interested in the modification of the lightcone for the perturbations.  The quartic Galileon gives the following contribution to the equation of motion: 
\be
\label{eq:quarticEOM}
(\Box \pi)^3 - 3 \Box \pi (\partial_\mu \partial_\nu \pi)^2 + 2 (\partial_\mu \partial_\nu \pi)^3 \;.
\ee
 The first term does not change the lightcone aperture. As the linear equation of motion is $\Box \pi_0 = 0$, in the second term we are forced to put the two $\partial_\mu \partial_\nu \pi$ legs on the background, so that also this term does not change the speed of the fluctuations.
The third term evaluated on a static background gives
\be 
( \partial_i \partial_j \pi_0 \partial_j \partial_k \pi_0)\,  \partial_k \partial_i \pi \,,
\ee       
and the matrix in brackets is positive definite. This implies that the coefficient of this operator must be negative to give subluminal propagation of perturbations.
\end{enumerate}

Let us see whether these conditions are compatible with the existence of a NEC violating solution. The equation of motion for a `de Sitter' solution can be easily derived from \cite{Nicolis:2008in} using the following trick. For a time dependent solution the term $(\partial\pi)^2\Box\pi$ does not contribute to the equations of motion, so that the results of  \cite{Nicolis:2008in} can be used simply replacing the coefficient of the cubic conformal Galileon $c_3$ with the one of $(\partial\pi)^4$ (\footnote{In this Appendix the coefficient of the $(\partial\pi)^4$ operator is given by $-\frac14 \frac{f^2}{H_0^2} \alpha (\partial\pi)^4$: in this way $\alpha$ simply replaces $c_3$ in the equations of motion.}). This gives
\be
-2 c_2 + 3 \alpha H_0^2 - 3 c_4 H_0^4 +\frac32 c_5 H_0^6= 0 \;.
\ee
Also the arguments of  \cite{NRT} to calculate the stress-energy tensor can be used; indeed the cubic Galileon is anyway treated separately. We get
\be
p = \frac13 T^\mu_\mu = - \frac{1}{H_0^2 t^4} \left(-c_2 +\alpha H_0^2 +\frac92 c_4 H_0^4 -\frac32 c_5 H_0^6 \right) =  - \frac{1}{H_0^2 t^4} \left(-3 c_2 + 4 \alpha H_0^2 +\frac32 c_4 H_0^6\right) \;,
\ee
where in the last step we used the equations of motion above.
The quantity in brackets should be positive to violate the NEC, but all the terms are constrained to be negative because of the conditions discussed above. This shows the impossibility to get an healthy Minkowski background, even before imposing the additional constraint from the stability of the NEC violating solution we found.

The previous discussion brings up a possible loophole in the arguments of the paper. To have an healthy NEC-violating solution, we considered the breaking of $SO(4,2)$. But would it be possible to achieve the same result, preserving the original symmetry, by simply requiring that the coefficient of the cubic Galileon vanishes around the NEC-violating solution and that the quartic Galileon has the healthy negative sign discussed above? Following the recipe of \cite{Nicolis:2008in,NRT} it is straightforward\footnote{The equation of motion and the condition $d_3 = 0$ (following the notation of \cite{Nicolis:2008in,NRT}) read respectively 
\be
-2 c_2 + 3 c_3 H_0^2 - 3 c_4 H_0^4 +\frac32 c_5 H_0^6= 0 \qquad c_3 - 3 c_4 H_0^2 + 3 c_5 H_0^4= 0 \;.
\ee
With these we can solve for $c_4$ and $c_5$ in terms of $c_2$ and $c_3$. Then we can impose the following three inequalities: $p <0$ (i.e.~NEC-violation), $d_2 >0$ and $d_4<0$ that read respectively
\be
-5 c_2 + \frac{11}{2} c_3 H_0^2 >0 \qquad - c_2 + \frac{1}{2} c_3 H_0^2 >0 \qquad \frac43 c_2 - c_3 H_0^2 < 0 \;.
\ee
Clearly these conditions are all compatible. 
} to check that this is indeed possible (though not compatible, as in the rest of the paper, with an healthy Minkowski vacuum). However, one quickly realizes that the situation is not completely satisfactory. A negative coefficient of the quartic Galileon ensures the subluminality of perturbations around time-independent solutions. If this may be a good criterion around Minkowski with non-relativistic sources, it is clearly not enough for the NEC-violating time-dependent solution, where space- and time-dependent perturbations must be considered. The last term of \eqref{eq:quarticEOM}, however, induces also superluminal corrections to the lightcone once time-dependent backgrounds are considered.

\end{document}